\documentclass[aps,prl,twocolumn,superscriptaddress]{revtex4-1}
\usepackage{amsmath}
\usepackage{graphicx}
\usepackage{color}
\usepackage{gensymb}
\usepackage{hyperref}
\usepackage{upgreek}
\usepackage{braket}

\newcommand{\micro}{${\upmu}$}

\begin{document}

\title{Valley coherent exciton-polaritons in a monolayer semiconductor}

\author{S. Dufferwiel}
\email{s.dufferwiel@sheffield.ac.uk}
\author{T. P. Lyons}
\email{tplyons1@sheffield.ac.uk}
\affiliation{Department of Physics and Astronomy, University of Sheffield, Sheffield S3 7RH, UK}
\author{D. D. Solnyshkov}
\affiliation{Institut Pascal, University Clermont Auvergne and CNRS, 24 Avenue Blaise Pascal, 63178 Aubiere Cedex, France}
\author{A. A. P. Trichet}
\affiliation{Department of Materials, University of Oxford, Parks Road, Oxford OX1 3PH, UK}
\author{F. Withers}
\affiliation{Centre for Graphene Science, CEMPS, University of Exeter, Exeter, EX4 4QF, UK}
\author{G. Malpuech}
\affiliation{Institut Pascal, University Clermont Auvergne and CNRS, 24 Avenue Blaise Pascal, 63178 Aubiere Cedex, France}
\author{J. M. Smith}
\affiliation{Department of Materials, University of Oxford, Parks Road, Oxford OX1 3PH, UK}
\author{K. S. Novoselov}
\affiliation{School of Physics and Astronomy, University of Manchester, Manchester M13 9PL, UK}
\author{M. S. Skolnick}
\author{D. N. Krizhanovskii}
\affiliation{Department of Physics and Astronomy, University of Sheffield, Sheffield S3 7RH, UK}
\author{A. I. Tartakovskii}
\email{a.tartakovskii@sheffield.ac.uk}
\affiliation{Department of Physics and Astronomy, University of Sheffield, Sheffield S3 7RH, UK}
\date{\today}

\begin{abstract}

Two-dimensional transition metal dichalcogenide (TMD) semiconductors provide a unique possibility to access the electronic valley degree of freedom using polarized light, opening the way to valley information transfer between distant systems. Excitons with a well-defined valley index (or valley pseudospin) as well as superpositions of the exciton valley states can be created with light having circular and linear polarization, respectively. However, the generated excitons have short lifetimes (ps) and are also subject to the electron-hole exchange interaction leading to fast relaxation of the valley pseudospin and coherence. Here we show that control of these processes can be gained by embedding a monolayer of WSe$_2$ in an optical microcavity, where part-light-part-matter exciton-polaritons are formed in the strong light-matter coupling regime. We demonstrate the optical initialization of the valley coherent polariton populations, exhibiting luminescence with a linear polarization degree up to 3 times higher than that of the excitons. We further control the evolution of the polariton valley coherence using a Faraday magnetic field to rotate the valley pseudospin by an angle defined by the exciton-cavity-mode detuning, which exceeds the rotation angle in the bare exciton. This work provides unique insight into the decoherence mechanisms in TMDs and demonstrates the potential for engineering the valley pseudospin dynamics in monolayer semiconductors embedded in photonic structures.

\end{abstract}


\maketitle

\begin{figure}
\includegraphics[scale=1]{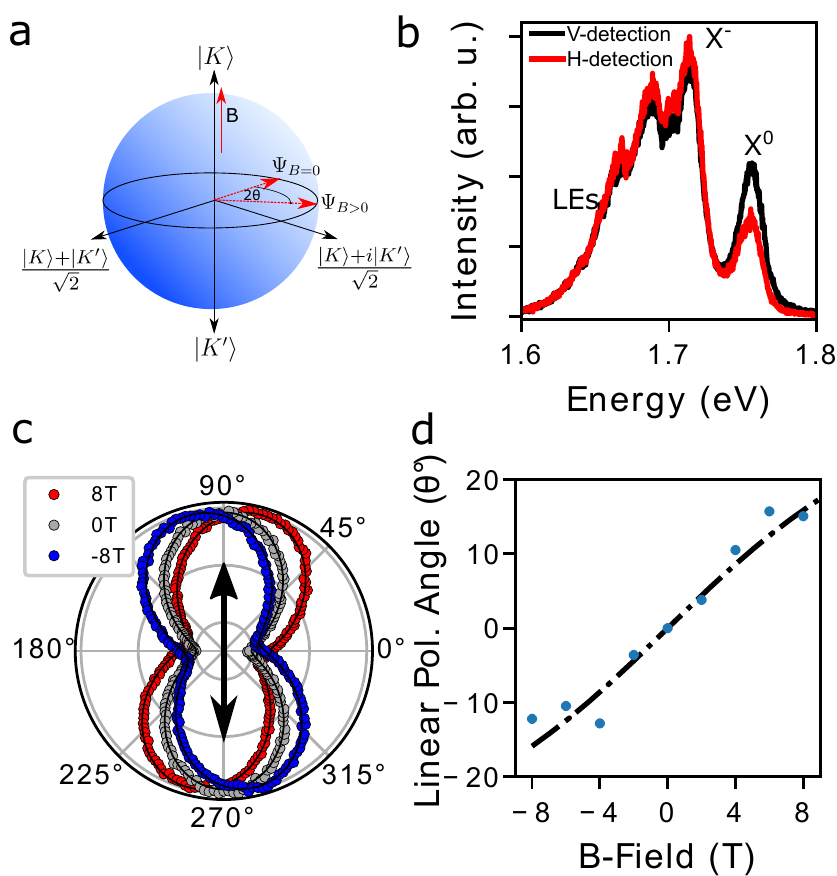}
\caption{\label{fig:fig1} \textbf{Valley coherence in WSe$_2$ excitons.}
\textbf{a} Bloch sphere representation of the valley pseudospin vector. Valley polarized states ($\Ket{K}$ and $\Ket{K'}$) lie on the poles of the sphere while valley coherence is represented by a Bloch vector oriented on the equator. The application of a Faraday field leads to precession of the pseudospin vector around the equator due to the valley Zeeman effect, evolving from a position $\Psi_{B=0}$ to $\Psi_{B>0}$. \textbf{b} Polarization resolved spectra under linear vertical excitation. Retention of injected valley coherence is present for the neutral exciton X$^0$ with a polarization degree of 15\%.  \textbf{c} Rotation of the linear polarization vector around the equatorial plane through the application of Faraday fields of $8$ T, $0$ T and $-8$ T respectively. The black arrow indicates the injected linear polarization. \textbf{d} The linear polarization angle ($\theta$) as a function of applied field. The fit corresponds to $\theta = \arctan{(\Omega_BT_{2})}/2$ where $\Omega_B$ is the precession frequency given by $\Omega_B = g\mu_B B /\hbar$ and $T_2$ is the coherence time. The fit corresponds to $T_2 = 0.52\pm0.05$ ps.}
\end{figure}

In monolayers of semiconducting transition metal dichalcogenides (TMD) inversion symmetry breaking, strong spin-orbit coupling and time reversal symmetry lead to the locking of the electronic spin orientation to the specific valley, $K$ or $K'$, at the edge of the Brillouin zone \cite{Xiao2012,Mak2012,Xu2014}. This observation has led to renewed interest in valleytronics with proposals to use monolayers of semiconducting TMDs for encoding information in their electronic valley degree of freedom, similar to the approach adopted for spins in spintronics. As with the formalism used for spins, the evolution of the pseudospin can be depicted on a Bloch sphere \cite{Xiao2012,Mak2012,Wang2016}, as shown in Fig.~\ref{fig:fig1}a, where the poles correspond to states $\ket{K}$ and $\ket{K'}$ with a well defined valley index, and the equatorial plane corresponds to a linear superposition of these states. 

It has been expected that owing to the large spin-orbit splittings in TMDs, the valley pseudospin will be robust against intervalley scattering \cite{Xiao2012,Mak2012}. Indeed, optical initialization of the valley states ($\ket{K}$ and $\ket{K'}$) by circularly polarized light has been widely observed for excitons in TMDs such as MoS$_2$, WS$_2$ and WSe$_2$ \cite{Sallen2012,Mak2012,SuzukiR2014,Zhu2014,Surrente2017} as well as for exciton-polaritons in MoSe$_2$ and MoS$_2$ embedded in optical cavities \cite{Chen2017,Dufferwiel2017,Sun2017,Lundt2017}. 
In addition, retention of linear polarization has been reported in WSe$_2$ and WS$_2$, indicating optical initialization of a superposition of valley states $\ket{X} = \frac{1}{\sqrt{2}}(\ket{K}+\ket{K'})$ \cite{Zhu2014,Wang2016,Jones2013,Wang2015,Hao2016,Schmidt2016,Ye2017} (see Fig.~\ref{fig:fig1}a). However, the lifetime of such exciton valley coherence has been estimated to be a few hundred femtoseconds, limiting the coherent manipulation of the valley pseudospin \cite{Wang2016,Hao2016,Schmidt2016,Ye2017}. The dephasing has been linked to the effect of the excitonic longitudinal-transverse (LT) splitting \cite{Wang2016,Hao2016,Schmidt2016,Ye2017}, arising from the electron-hole exchange interaction \cite{Yu2014_1,Yu2014_2}, which can be viewed as an effective magnetic field that acts on the exciton valley pseudospin. This splitting reaches a few meV and is proportional to the magnitude of the exciton momentum $k$, thus strongly affecting excitons generated in high energy states by non-resonant optical excitation. The scattering of the exciton k-vector due to disorder creates a randomly varying effective magnetic field leading to random pseudospin precession and associated valley depolarization, and is known as the Maialle--Silva--Sham (MSS) mechanism \cite{Maialle1993}. Another factor limiting the ability to manipulate the valley coherence is the short exciton lifetime (1-2 ps)\cite{Wang2016,Hao2016,Schmidt2016,Ye2017}, which in principle can be overcome by using the valley  index of electrons or holes, with the disadvantage of reduced optical control \cite{Yang2015,Kim2017}.  

\begin{figure}
\includegraphics{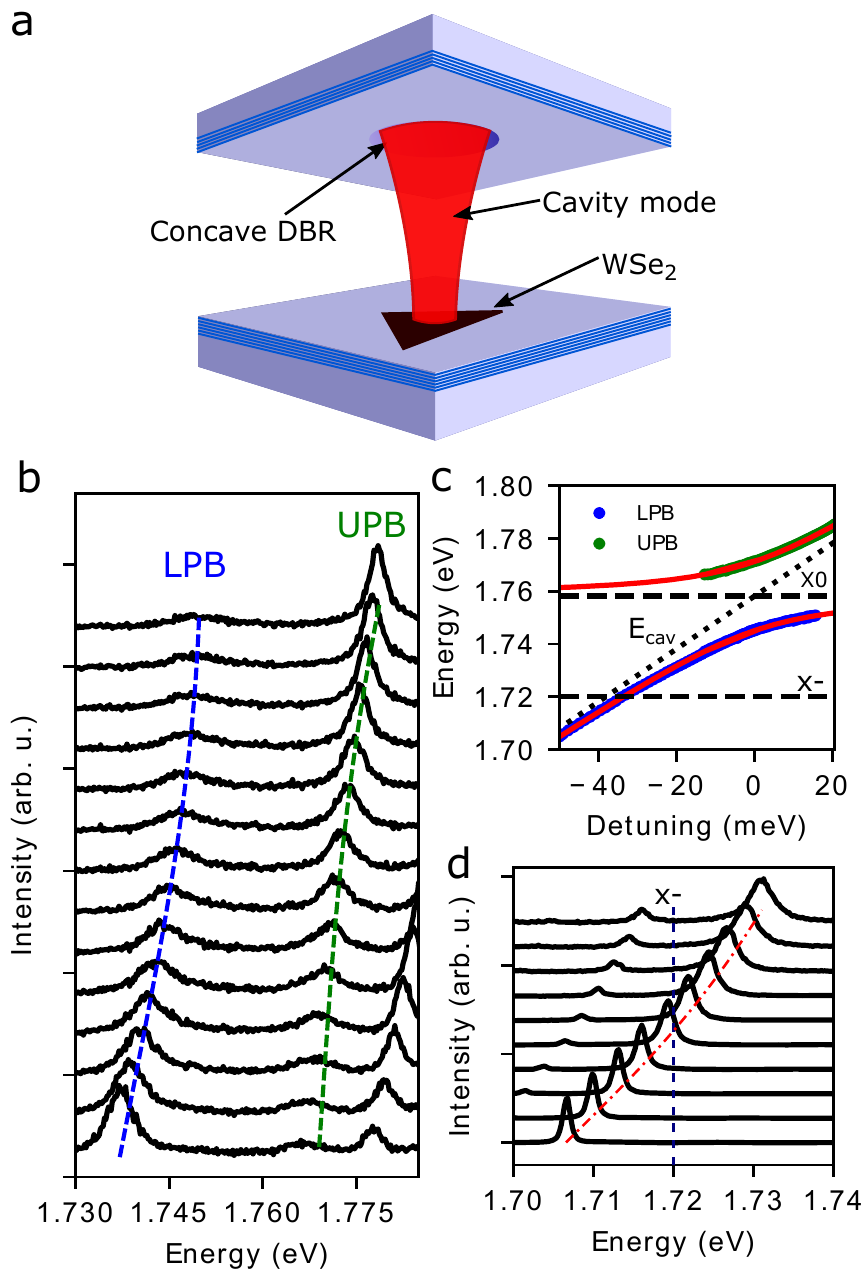}
\caption{\label{fig:fig2} \textbf{Strong exciton-photon coupling in WSe$_2$ monolayers.}
\textbf{a} Schematic of a WSe$_2$ monolayer placed in an open hemispherical microcavity. \textbf{b} Photoluminescence spectra measured as the cavity mode is scanned across the exciton resonance. The graph shows formation of the lower and upper polariton branches, LPB and UPB, respectively. \textbf{c}  Extracted polariton peak positions as a function of the exciton-photon detuning ($\Delta = E_c - E_{X^0}$), where $E_c$ and $E_{X^0}$ are the tunable cavity and exciton energies. The green (blue) symbols show spectral positions of the UPB (LPB) peaks. The red lines correspond to a coupled oscillator model fit with a Rabi splitting of $26.2 \pm 0.1$ meV. \textbf{d} Spectra corresponding to tuning the ground state mode through the trion energy demonstrating weak coupling between the cavity and trion.}
\end{figure}

Recently, an alternative approach enabling new ways to control the valley pseudospin has emerged in experiments on monolayer MoSe$_2$ and MoS$_2$ embedded in optical microcavities \cite{Chen2017,Dufferwiel2017,Sun2017,Lundt2017}, where part-light-part-matter exciton-polaritons are formed due to the strong coupling of the cavity mode and excitons with $k\approx 0$. The polaritons exhibit a modified energy spectrum with upper and lower polariton branches split by a few tens of meV (see Fig.~\ref{fig:fig2}) and are significantly less sensitive to disorder compared with the excitons. Both effects lead to modified valley pseudospin dynamics resulting in an increase in retention of valley polarization in the polariton states \cite{Chen2017,Dufferwiel2017,Sun2017,Lundt2017}. In high-finesse tunable microcavities, as in the present work, additional control of the valley pseudospin dynamics can be gained by modifying the cavity-exciton energy detuning, which changes the exciton and photon fractions of the polariton states, thus influencing the polariton radiative and valley depolarization times as well as exciton energy relaxation \cite{Dufferwiel2017}.

Here we show that by embedding WSe$_2$ in an optical cavity in the strong light-matter coupling regime we can access valley coherent exciton-polariton populations and control their dynamics by varying the exciton-cavity-mode detuning and applied magnetic field. These quasiparticles are composed of excitons in a coherent linear superposition of valley states which are in a further superposition of exciton and photon states. We demonstrate that strong suppression of valley dephasing can be achieved for polaritons resulting in a threefold enhancement of linear polarization degree observed for the upper polariton branch (UPB). On the other hand, the lower polariton branch (LPB) shows a linear polarization degree that is strongly dependent on the exciton-photon detuning. A dynamical model, incorporating cavity-modified exciton relaxation, detuning-dependent polariton lifetimes as well as disorder scattering in the presence of the excitonic LT-splitting, reproduces the exciton-cavity detuning dependence of the linear polarization degree. The model and experiments confirm the exciton LT-splitting as the dominant mechanism for exciton dephasing. We then demonstrate coherent manipulation of the polariton valley pseudospin through the application of a magnetic field in the Faraday geometry, which induces precession of the valley pseudospin around the equator of the Bloch sphere due to the valley Zeeman effect. This rotation of the valley pseudospin vector for the LPB is up to 3 times larger than the bare exciton and can be accurately varied through selection of the exciton-photon detuning where the final linear polarization angle is determined by the amount of pseudospin precession of high k-vector reservoir excitons during relaxation and and the smaller pseudospin precession of polariton states before radiative decay. The generation and control of valley coherence in WSe$_2$ exciton-polaritons creates new opportunities for manipulation of the valley pseudospin in TMD monolayers embedded in optical cavities. This work also opens the way to unexplored non-linear polariton phenomena utilizing the valley degree of freedom in TMDs in polariton condensates, the optical spin Hall effect, optical spin switching and polarization bistabilities \cite{Carusotto2013}.

\begin{figure*}
\includegraphics{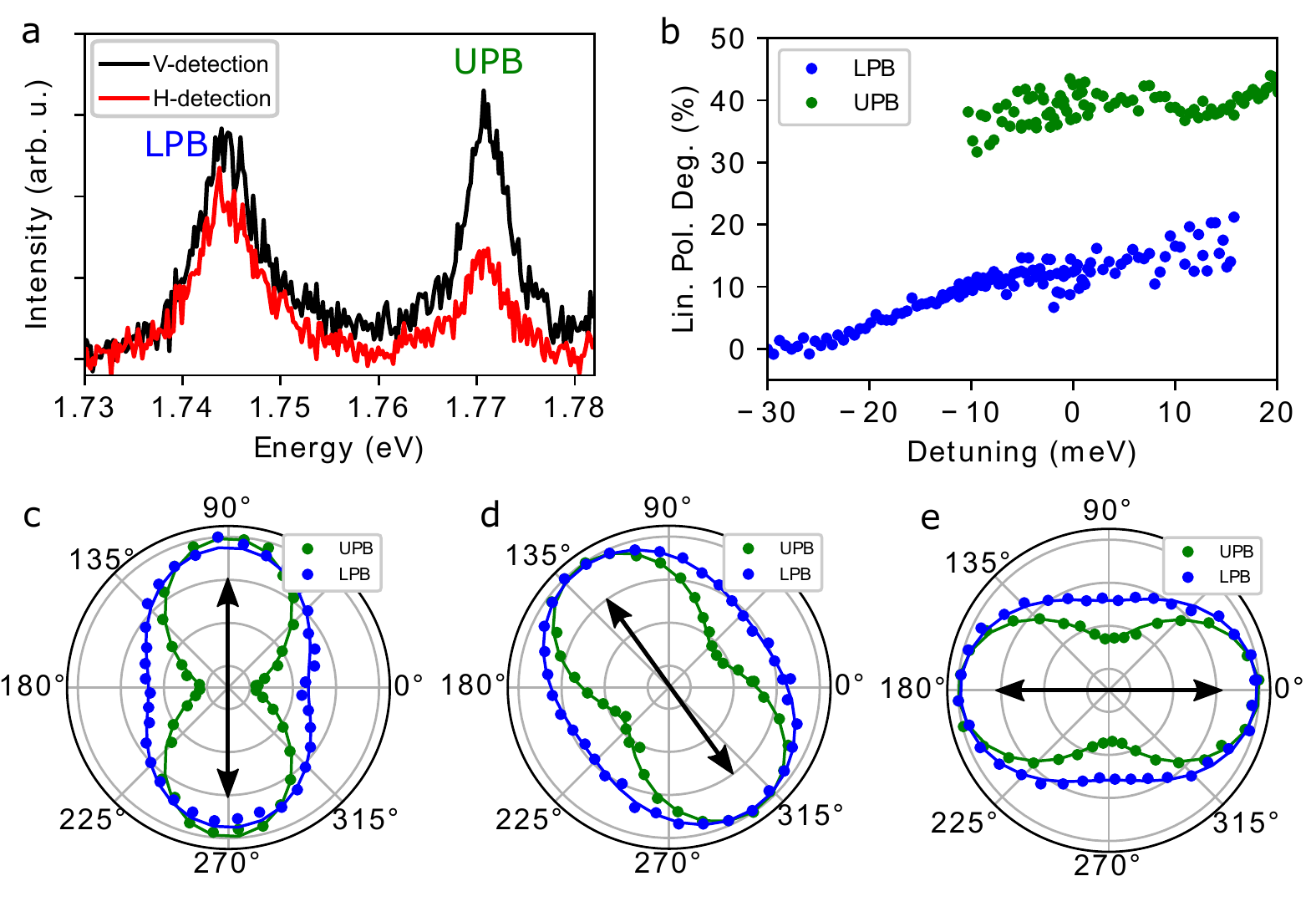}
\caption{\label{fig:fig3} \textbf{Valley coherence of exciton-polaritons in WSe$_2$.}
\textbf{a} Polarization resolved spectra under vertical linear excitation showing clear retention of injected valley coherence in the polariton branches. \textbf{b} Linear polarization degree as a function of exciton-photon detuning under vertically linearly polarized excitation for both the LPB (blue) and UPB (green). \textbf{c,d,e} Angular dependent intensity polar plots for vertical (\textbf{c}), diagonal (\textbf{d}) and horizontal (\textbf{e}) excitation (black arrows) at zero exciton-photon detuning. The data for LPB (UPB) are shown with blue (green) symbols.}
\end{figure*}

\section{Results}

The WSe$_2$ sample consists of a single monolayer placed at the surface of a planar distributed Bragg reflector (DBR). An optical image of the flake is shown in Supplementary Fig. S1. Characterization of the sample was performed under non-resonant excitation at $1.946$ eV at $4.2$ K and spectra are shown in Fig.~\ref{fig:fig1}\textbf{b}. Features associated with a neutral exciton ($X^0$) and charged exciton ($X^-$) can be identified along with a large band of localized emitters (LEs) at lower energy. Under linear vertical excitation (Fig.~\ref{fig:fig1}\textbf{b}) the exciton resonance shows retention of linear polarization with a polarization degree of around 15\%, indicating the optical initialization of a coherent superposition of valleys. The lack of linear retention for the trion peak has been attributed to the trion fine structure which leads to rapid dephasing of trion coherence \cite{Jones2013}. In this sample, a small negative polarization degree of $-2$\% is observed for the trion and is not attributed to valley coherence. Clear retention of valley polarization under circular excitation is also present for both excitonic species with a circular polarization degree of around 30\% and 50\% for the exciton and trion respectively as shown in Supplementary Fig. S2.

In order to demonstrate control of the valley coherent exciton states we apply a magnetic field in the Faraday geometry \cite{Wang2016,Schmidt2016}. This lifts the degeneracy of the exciton valley states $\ket{K}$ and $\ket{K'}$ due to the valley Zeeman effect in WSe$_2$\cite{Srivastava2015,Aivazian2015}, with the splitting given by $\hbar\Omega_B = g\mu_BB$. Here, $g$ is the exciton g-factor, measured to be 1.7 for this sample (see Supplementary Fig. S3). After initialization, the linear valley superposition state will evolve with time asy $\ket{X} = \frac{1}{\sqrt{2}}(\ket{K}e^{-i\Omega_B t/2} + \ket{K'}e^{i\Omega_B t/2})$. As such, after a certain time the exciton ensemble state will evolve to a new position on the equator of the Bloch sphere with a corresponding rotation of the linear polarization angle. Fig.~\ref{fig:fig1}\textbf{c} shows the photoluminescence (PL) intensity as a function of linear polarization angle for applied fields of $-8$ T, $0$ T and $+8$ T under vertical linear excitation, where clear rotation of the valley coherent pseudospin in PL can be seen. The extracted rotation angle as a function of magnetic field is shown in Fig.~\ref{fig:fig1}\textbf{d}. The fit corresponds to $\theta = \arctan(\Omega_BT_2)/2$ where $T_2$ is the fitted coherence time, defined as $1/T_2=1/2T_1 + 1/T_2^*$ where $T_1$ and $T_2^*$ are the state lifetime and pure dephasing times respectively. $T_2$ is extracted to be $0.52 \pm 0.05$ ps, consistent with previous reports \cite{Hao2016,Wang2016,Schmidt2016}.

The tunable optical microcavity is formed by introducing a top concave DBR into the optical path using piezo nanopositioners and bringing the two mirrors to a total optical cavity length of around 2.5 \micro m. A schematic diagram with embedded monolayer is shown in Fig.~\ref{fig:fig2}\textbf{a}. The formed microcavity is hemispherical and supports zero-dimensional Laguerre-Gaussian modes. In this work only the coupling with the ground state longitudinal mode is discussed. By changing the top and bottom mirror separation, the cavity resonance can be scanned across the exciton resonance. In this experiment we observe a characteristic anti-crossing signifying the formation of exciton-polariton branches, as shown in Fig.~\ref{fig:fig2}\textbf{b} for the longitudinal mode. Fitting the peak positions with a coupled oscillator model yields a Rabi splitting of $26.2\pm0.1$ meV (Fig.~\ref{fig:fig2}\textbf{c}). Due to the low intrinsic electron doping present in the sample, weak coupling is observed between the trion and cavity \cite{Dufferwiel2017,Sidler2016,Dufferwiel2015}, as shown in Fig.~\ref{fig:fig2}\textbf{d}.

To probe retention of valley coherence in the polaritonic system we excite non-resonantly at $1.946$ eV with linearly polarized light at zero exciton-cavity-mode detuning. Fig.~\ref{fig:fig3}\textbf{a} shows the polarization resolved spectra co- and cross-polarized to excitation. It is clear that both the LPB and UPB show retention of the linear polarization and hence valley coherence. Interestingly, the UPB shows a much larger degree of polarization of around $40$\% in contrast to the LPB value of around 15\%, which is comparable to the bare exciton. 

In-situ tunability of the cavity mode energy allows the degree of linear polarization to be probed as a function of exciton-photon detuning $\Delta = E_c-E_{X^0}$, where $E_c$ and $E_{X^0}$ are the cavity  and exciton energies respectively. The result is plotted in Fig.~\ref{fig:fig3}\textbf{b}. At large negative detuning the LPB polarization degree is close to zero and increases as the exciton resonance is approached. The maximum polarization degree of around 15\% is measured when the LPB is highly excitonic and the value reflects the bare exciton polarization degree. In contrast, the UPB shows up to a $3$ times enhancement in polarization degree relative to the bare flake. Moreover, an increase in the UPB polarization degree of around $10\%$ is observed as the detuning is changed from $-10$ meV to $+25$ meV. In order to confirm that the valley coherence is inherited from the non-resonant excitation we rotate the orientation of the linearly polarized pump and record the angular dependent intensity in collection. The resultant polar plots are shown in Fig.~\ref{fig:fig3}\textbf{c}, \textbf{d} and \textbf{e} for vertical, diagonal and horizontal excitation respectively. Clear retention of the injected valley coherence is demonstrated. Strong retention of valley polarization is also observed under circularly polarized excitation and detection as shown in Supplementary Fig. S4.

\begin{figure}
\includegraphics{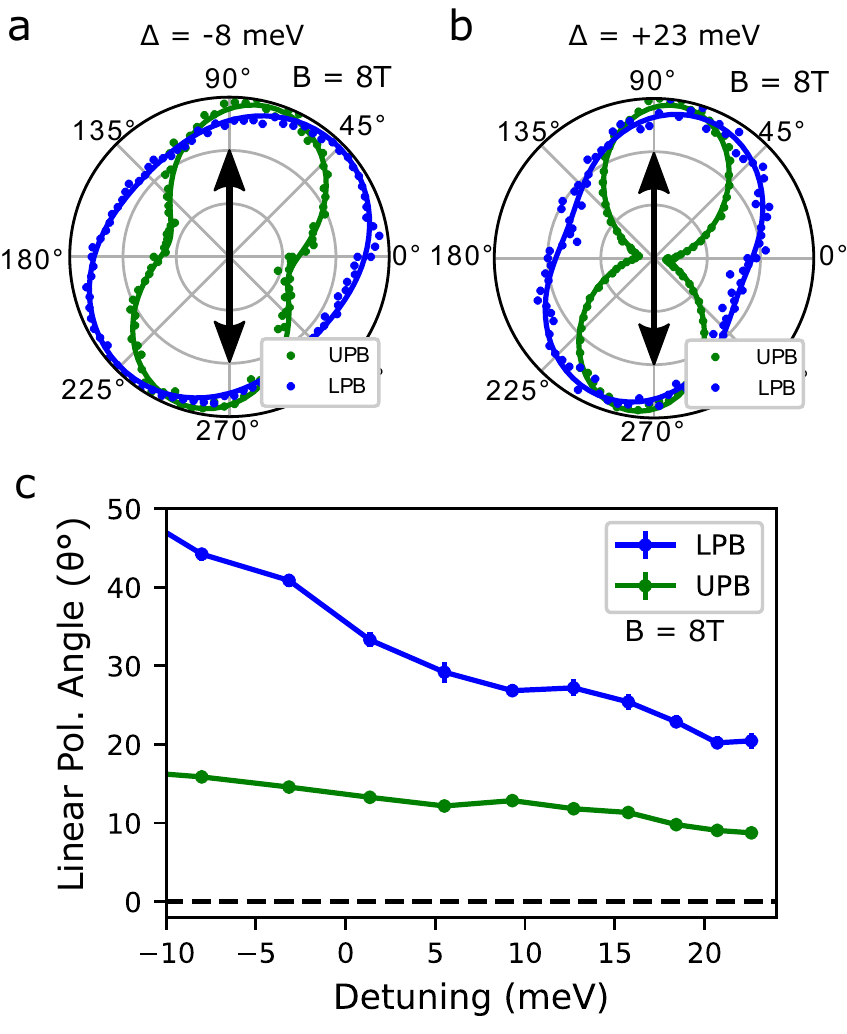}
\caption{\label{fig:fig4} \textbf{Control of the exciton-polariton valley pseudospin vector in a Faraday magnetic field.} \textbf{a,b} Polariton intensity as a function of the varied polarization angle in detection under vertically linearly polarized excitation and a Faraday field of $B=8$ T for detunings of \textbf{a} $\Delta=-8$ meV  and \textbf{b} $\Delta= +23$ meV. The black arrows indicate the injected valley coherence (polarization of the excitation laser). \textbf{c} The polarization rotation angle in polariton PL as a function of the exciton-photon detuning at $B=8$ T. The data for LPB (UPB) are shown in blue (green).}
\end{figure}

To demonstrate control of the valley coherent polariton population we apply a magnetic field in the Faraday geometry as discussed previously for the exciton. Fig.~\ref{fig:fig4}\textbf{a} and \textbf{b} show the polariton PL as a function of polarization angle under vertical excitation and at exciton-photon detunings of $-8$ meV and $+23$ meV respectively for an applied magnetic field of $B = 8$ T. It is clear from the plots that the induced rotation of the LPB is much larger than that of the UPB, the latter comparable to the bare exciton rotation. In order to probe the effect of the exciton-photon detuning we plot the rotation angle of the LPB and UPB at a fixed field of $B = 8$ T and sweep the detuning from $-10$ meV to $+25$ meV. The resultant rotation angles are plotted in Fig.~\ref{fig:fig4}\textbf{c} where a clear increase in the rotation angle is present for increasing negative detuning. Significantly, the LPB rotation approaches $50^\circ$ at $-10$ meV, a factor of almost 3 times larger than the bare exciton. At detunings below $-10$ meV the weak coupling of the LPB to the trion resonance masks the probing of the coherent rotation due to the small orthogonal polarization degree of the bare trion resonance, which is emitted by the near-resonant LPB as shown in Supplementary Fig. S5. Significantly, the controlled selection of the exciton-photon detuning allows the arbitrary selection of the degree of rotation of the polarization plane for a given field, introducing a new degree of control of the valley pseudospin. While magnetic fields have been used in this work, optical pulses as used in \cite{Ye2017} could be combined with a selected exciton-photon detuning to allow enhanced and high speed rotation of the valley pseudospin vector.

\begin{figure}
\includegraphics{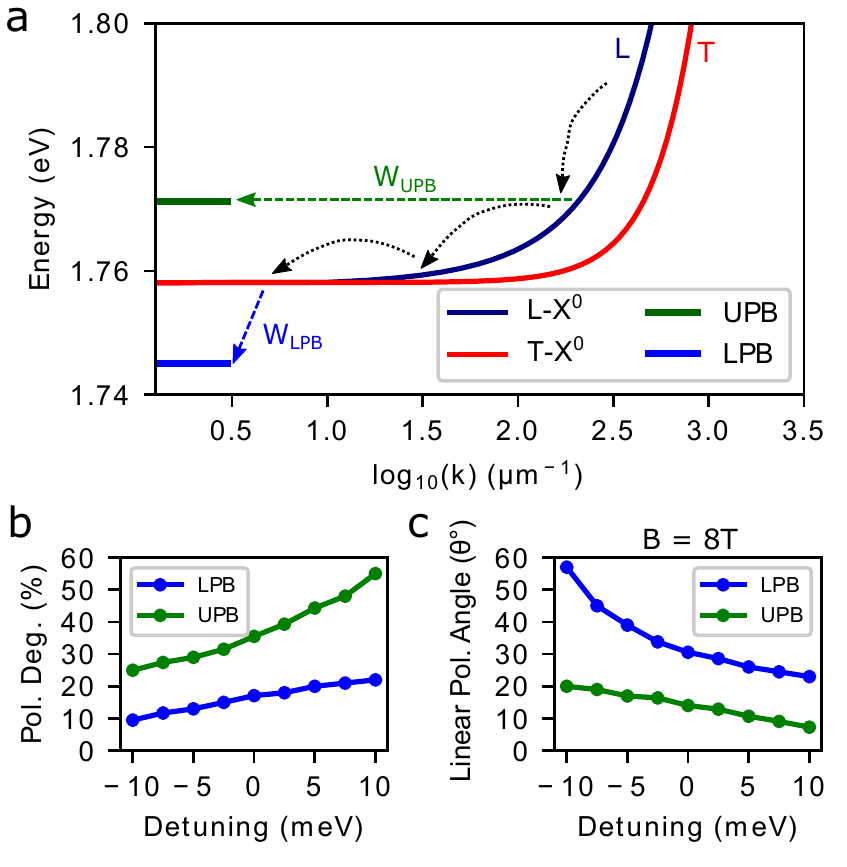}
\caption{\label{fig:fig5} \textbf{Theoretical model for relaxation of valley coherent excitons and polaritons.} \textbf{a} Dispersion of polariton and LT-split exciton states (L-X$^0$, T-X$^0$) showing the theoretical model of valley coherent exciton relaxation where dashed and dotted lines indicate relaxation pathways of excitons with $W_{UPB}$ and $W_{LPB}$ describing the corresponding relaxation rates. \textbf{b} Simulated linear polarization degree of the polariton branches as a function of detuning. \textbf{c} Simulated rotation angle of the linear polarization in the polariton branches with an applied Faraday magnetic field of $8$ T as a function of the exciton-photon detuning. The data for LPB (UPB) are shown in blue (green).}
\end{figure}

To describe the system and the resultant valley coherent polariton population we consider the dynamical relaxation processes outlined in Fig.~\ref{fig:fig5}\textbf{a} where initially we have no applied external magnetic field. We optically excite valley coherent carriers which scatter to form high in-plane wavevector (k-vector) excitons at the pump energy of 1.946 eV which relax quickly along the exciton dispersion due to phonon scattering to low k-vectors. Here, these reservoir excitons are influenced by the effective spin-orbit coupling which arises from the exciton LT-splitting and results in an effective magnetic field acting on the valley pseudospin \cite{Yu2014_1,Yu2014_2,Maialle1993,Glazov2014}. The magnitude and direction of the effective field depends on the exciton propagation direction  \cite{Yu2014_1,Yu2014_2,Maialle1993,Glazov2014,Kavokin2005}. Scattering of these excitons due to disorder creates a randomly varying field which leads to random valley pseudospin precession and the associated loss of the valley coherence \cite{Wang2016,Hao2016,Schmidt2016,Ye2017}. Because of this dephasing process, determined by the LT splitting and disorder, both the degree of linear polarization and the rotation angle observed in the polariton luminescence are dependent on the time the excitons spend in the reservoir. The polarization degree is higher for excitons relaxing to the polariton states quickly, while the rotation angle is higher in the polariton states populated by excitons which spend more time in the reservoir.  

The UPB is populated through direct disorder scattering of excitons which are degenerate with the polariton branch due to the parabolic exciton dispersion. The LPB is populated by relaxation of excitons from the bottom of the exciton reservoir. This scattering rate is strongly dependent on the exciton-photon detuning where at large negative detunings the scattering is strongly suppressed due to the large energy mismatch. This leads to accumulation of excitons in the reservoir, and their slow relaxation to the LPB. Once relaxed to the polariton branches, radiative decay occurs due to the polariton photonic component where the linear polarization degree quantifies the level of preservation of coherence during the relaxation process. The derived rate equations can be found in Supplementary Note 1. 

Fig.~\ref{fig:fig5}\textbf{b} shows the simulated linear polarization degree as a function of detuning in the absence of an applied magnetic field as measured experimentally in Fig.~\ref{fig:fig3}\textbf{b}. A full list of used parameters can be found in Supplementary Note 1. Both polariton branches exhibit the experimentally observed increase in linear polarization degree when moving towards positive detuning. The qualitative dependence for the LPB is explained by the enhancement of the coupling of the exciton reservoir with the LPB. The exciton scattering rate into the LPB states increases with positive detuning because of their increased exciton fraction. Thus, the excitons populating these LPB states spend less time in the reservoir, and therefore have a shorter time to lose their polarization degree. The UPB, because of the resonant coupling to high k-vector excitons, directly probes the linear polarization degree of the reservoir states. The higher the energy probed, the higher the polarization degree, because loss of polarization degree accompanies the energy relaxation of the excitons. In addition, as the photonic fraction of the UPB increases the net scattering rate from the exciton reservoir to the UPB increases as well, since the higher photonic fraction ensures that UPB polaritons radiatively decay, this inhibits scattering back to the exciton reservoir where strong depolarization occurs, and hence a higher polarization degree is expected.

Next, we consider the application of a Faraday field,  which leads to precession of the pseudospin in both reservoir excitons and polaritons (details can be found in Supplementary Note 1). Fig.~\ref{fig:fig5}\textbf{c} shows the simulated rotation angle of the linear polarization plane under the effect of an applied magnetic field of $B=8$ T. As observed experimentally, both branches exhibit a decrease in rotation angle with increasing detuning, and the rotation of the UPB is always smaller than the LPB. Qualitatively, this behaviour is again explained by the time spent by the excitons in the reservoir as for the detuning dependence in Fig.~\ref{fig:fig3}\textbf{b}. Here, the longer the time spent in the reservoir, the higher the rotation angle. Some rotation of the valley pseudospin also occurs in the polariton branches, but the lower exciton component leads to reduced Zeeman splitting and precession frequencies, and hence a smaller effect than the rotation which occurs in the reservoir. For the LPB, positive detuning means better coupling to the reservoir, and so excitons relax quickly and have shorter time to rotate their linear polarization before relaxing into the polariton state. The UPB is populated from the high energy high k-vector exciton states. Excitons have rapidly relaxed to these energies, and their polarization plane has not yet rotated strongly before they experience fast momentum scattering and populate the resonant UPB states. As the positive detuning is increased, the UPB states probe excitons with progressively higher energies, who spend less time in the reservoir, and so polarization rotation observed in the UPB states becomes smaller. 
 
In conclusion, we report the generation and control of valley coherent populations of exciton-polaritons in a WSe$_2$ monolayer embedded in a microcavity. Our phenomenological model together with the experimental measurements of the polarized polariton luminescence shed new light on exciton valley dynamics, and confirm that the exciton LT-splitting, which arises due to the long-range Coulomb exchange interaction, is the dominant mechanism of exciton dephasing. By utilizing the strong light-matter coupling regime, we probe the valley coherence dynamics in the otherwise dark exciton reservoir states in a wide range of energies up to 30 meV above the $k=0$ exciton states. This reveals that the exciton energy relaxation is accompanied with the decay of their valley coherence, and demonstrates that excitons with a high degree of valley coherence in the states resonant with the high energy upper polariton branch can be efficiently extracted from the reservoir. Further implications of the interplay between the exciton valley coherence and energy-momentum relaxation dynamics are revealed in the rotation of the linear polarization of polariton luminescence under a Faraday magnetic field. We show that the fundamentally different exciton relaxation pathways into the upper and lower polariton branches result in the ability to fully control the phase rotation of these robustly coherent states, simply via selection of the exciton-photon detuning, over a range which has a UPB related lower bound equal to the bare flake rotation, and an LPB related upper bound over three times larger.

Our work shows that by incorporating monolayer semiconducting TMDs in photonic structures, optical signatures of the excitonic valley coherence can be significantly modified and further controlled by the exciton and cavity frequency detuning. We anticipate that TMD-based polaritonic circuit elements may be used for valleytronic devices exploiting controlled dynamics of the polariton valley pseudospin as shown in this work. Such valley polaritonic systems allow for the possibility of extended spatial propagation in two-dimensional microcavities and waveguides as well as the potential for response to electric fields (particularly for trion-polaritons) with low sensitivity to disorder and, importantly, a strong non-linearity at high polariton densities. 

\subsection{Sample Preparation}

Monolayer sheets of WSe$_2$ were obtained through mechanical exfoliation of bulk crystals. A monolayer of WSe$_2$ was transferred onto the DBR surface using standard mechanical exfoliation and standard transfer techniques. Bulk crystals were acquired from HQGraphene.

\subsection{Optical Measurements}
Optical measurements were performed with samples held in a helium bath cryostat system at a temperature of 4.2 K. Top and bottom DBRs were attached to XYZ nanopositioners with additional goniometer stages allowing tilt control of the bottom DBR. Optical excitation of the bare monolayer was possible by removing the top DBR from the optical path. All \micro-PL experiments were performed with a continuous-wave (cw) excitation using a 638 nm laser diode, focused onto the sample with an achromatic lens. Polarization resolved measurements were performed using a combination of linear polarizer and a quarter waveplate in the excitation path, and quarter waveplate, half-wave plate and linear polarizer in the collection path, allowing linearly and circularly polarized excitation and detection. PL was collected by focusing onto a single mode fibre which was guided into a 0.75 m spectrometer and a high sensitivity charge-coupled device.

\subsection{Microcavity}

The tunable microcavity with embedded TMD monolayer is formed using an external concave mirror to produce a zero-dimensional tunable cavity \cite{Dufferwiel2014,Schwarz2014}. The formed cavity schematic is shown in Fig.~\ref{fig:fig2}\textbf{a} with the monolayer placed at an electric-field antinode, and nanopositioners are used to control the cavity spectral resonance energy. The nominal radius of curvature of the concave mirror is 20 \micro m leading to a beam waist on the planar mirror of around 1 \micro m \cite{Dufferwiel2014}.

\section{Acknowledgements}

The authors thank the financial support of the Graphene Flagship under grant agreement 696656, EPSRC grants EP/M012727/1, EP/P026850/1 and EP/J007544/1, ITN Spin-NANO under grant agreement 676108. A. A. P. T., D. N. K. and J. M. S. acknowledge support from the Leverholme Trust. K. S. N. thanks financial support from the Royal Society, EPSRC, US Army Research Office and ERC Grant Hetero2D.

\section{Author Contributions}
These authors contributed equally to this work: S. Dufferwiel and T. P. Lyons.

S.D. and T. P. L. carried out optical investigations. Sample fabrication was performed by F. W. and T. P. L. A. A. P. T. designed and fabricated the concave mirrors. D. D. S. and G. M carried out theoretical analysis. S. D. and T. P. L. analysed the data. S. D. and A. I. T. wrote the manuscript with contributions from all co-authors. J. M. S., K. S. N., M. S. S., D. N. K., and A. I. T. provided management of various aspects of the project. S.D. conceived the experiment.  A. I. T. oversaw the project.

%\bibliography{valley_coherence_wse2}

\newpage

%%%%%%%%%% Merge with supplemental materials %%%%%%%%%%
%%%%%%%%%% Prefix a "S" to all equations, figures, tables and reset the counter %%%%%%%%%%
\renewcommand{\thesection}{S\arabic{section}}
\setcounter{section}{0}
\renewcommand{\thefigure}{S\arabic{figure}}
\setcounter{figure}{0}
\renewcommand{\theequation}{S\arabic{equation}}
\setcounter{equation}{0}
\renewcommand{\thetable}{S\arabic{table}}
\setcounter{table}{0}
\renewcommand{\citenumfont}[1]{S#1}
\makeatletter
\renewcommand{\@biblabel}[1]{S#1.}
\makeatother

\onecolumngrid

\section{Supplementary Note 1: Theoretical model}

To describe the spin dynamics in the presence of various effective and real fields, we describe the exciton reservoir as a set of states corresponding to different energies $E$ and different propagation directions $\alpha$. Each state is described by the population $n(E,\alpha)$ and pseudospin $\mathbf{S}(E,\alpha)$. We assume that while the pseudospin distribution may indeed be different for different angles $\alpha$, populations are the same: $n=n(E)$. Two extra states describe the lower and upper polaritons, for which we also define the populations $n_{L/U}$ and the pseudospins $\mathbf{S}_{L/U}$. The pseudospin is normalized in such a way that its ratio to the number of particles gives a corresponding polarization degree: $\rho_i=S_i/n$.

The excitons in the reservoir are affected by the spin-orbit coupling, which results in an effective magnetic field determined by their propagation direction and velocity: 
\begin{equation}
	\mathbf{\Omega}_{SOC}(\mathbf{k})=\beta k(\cos 2\alpha,\sin 2\alpha,0)^T
\end{equation}
where $\alpha$ is the propagation direction and $\beta$ is the spin-orbit coupling constant.

The rate equations for the populations and the pseudospins of the reservoir states are written as:
\begin{equation}
	\begin{array}{c}
		\frac{d}{{dt}}n\left( E \right) = P\left( E \right) -\Gamma n\left( E \right) +\sum_{E'} W\left( E'-E  \right)n\left( {E' } \right)\\
		- \left( {W\left( { E-E' } \right) + {W_L}\left( E \right) + {W_U}\left( E \right)} \right)n\left( E \right)
	\end{array}
\end{equation}
where $P(E)$ describes pumping, $\Gamma$ is the non-radiative decay and the phonon-assisted scattering rates are given by:
\begin{eqnarray}
	W\left(E'-E>0\right)=W_{ph}\left(1+\exp\left(\frac{E-E'}{k_B T}\right)\right)\\
	W\left(E'-E<0\right)=W_{ph}\exp\left(\frac{E'-E}{k_B T}\right)
\end{eqnarray}
and the scattering rates towards the polariton branches are given by
\begin{equation}
	W_U\left(E\right)=W_{U0}\delta\left(E-E_{U}\right)
\end{equation}
and
\begin{equation}
	W_L\left(E\right)=W_{L0}x_L\exp\left(\frac{-(E-E_{L})^2}{2\sigma^2}\right)
\end{equation}
where $E_L$ and $E_U$ are the energies of the lower and upper polariton modes, $x_L$ is the excitonic fraction of the lower polariton. All these parameters depend on the detuning $\Delta$. Finally, $\sigma$ is the inhomogeneous broadening of the exciton line, which allows the excitons to scatter quite efficiently to the lower polariton mode, in spite of the fact that the latter is quite far. However, this scattering rate strongly decreases at negative detunings. The upper polariton is described as being resonantly populated from the reservoir mode which has the same energy, as ensured by the Dirac's delta function.

%\begin{figure}[tbp]
%\includegraphics[scale=0.35]{fig1.pdf}
%\caption{ (Color online) Linear polarization degree of the polariton modes (blue - LPB, green - UPB) as a function of the exciton-photon detuning.}
%\label{poldeglpb}
%\end{figure}

The rate equations for the pseudospin take into account not only the scattering between different energies, but also the rotation of the pseudospin around effective magnetic fields (spin-orbit coupling $\Omega_{SOC}$ and Zeeman field $\Omega_Z$) and the scattering between the different propagation directions caused by disorder.
\begin{equation}
	\begin{array}{c}
		\frac{d}{{dt}}\mathbf{S}\left( {E,\alpha } \right) = \mathbf{P}\left( E \right) + \mathbf{\Omega} \left( {E,\alpha } \right) \times \mathbf{S}\left( {E,\alpha } \right)\\
		+ {W_d}\left( E \right)\left( {\left\langle {\mathbf{S}\left( E \right)} \right\rangle  - \mathbf{S}\left( {E,\alpha } \right)} \right)\\
		+ \sum_{E'}W\left( {E' - E} \right)\mathbf{S}\left( {E',\alpha } \right)-\Gamma\mathbf{S}\left(E,\alpha\right)\\
		- \left( {W\left( {E - E'} \right) + {W_L}\left( E \right) + {W_U}\left( E \right)} \right)\mathbf{S}\left( {E,\alpha } \right)
	\end{array}
\end{equation}
Here, $\mathbf{\Omega}$ includes both the spin-orbit coupling $\mathbf{\Omega}(\mathbf{k})$ defined above and the Zeeman splitting $\Omega_Z$ if the latter is applied. The disorder-induced scattering $W_d$ couples all states with the same energy $E$. Its contribution tends to fill a state with a given direction $\alpha$ by particles having the spin equal to its average value $\langle \mathbf{S}(E)\rangle_\alpha$. The terms with $W$ are identical to those of the equations for the populations and keep the required normalization of the pseudospin: they do not affect the polarization degree and they cannot lead to the rotation of a pseudospin.

The rate equations for the populations of the polariton states read:
\begin{equation}
	\frac{d}{{dt}}{n_{L/U}} = \sum\limits_E {{W_{L/U}}\left( E \right)n\left( E \right)}  - \frac{{{n_{L/U}}}}{{{\tau _{L/U}}}}
\end{equation}
and for pseudospins:
\begin{equation}
	\frac{d}{{dt}}{{\bf{S}}_{L/U}} = \sum\limits_E {{W_{L/U}}\left( E \right)\left\langle {{\bf{S}}\left( E \right)} \right\rangle }  - \frac{{{{\bf{S}}_{L/U}}}}{{{\tau _{L/U}}}}+\mathbf{\Omega}_{L/U}\times\mathbf{S}_{L/U}
\end{equation}

In our simulations, we have used the following parameters. The spin-orbit coupling constant $\beta=40$ $\mu$eV$\cdot\mu m$ \cite{Dufferwiel2017}. The disorder-induced scattering $W_d=10^{13}$ s$^{-1}$ which determines the spin relaxation in the exciton reservoir together with the spin-orbit coupling. The value of the Zeeman splitting from the experiment (see Supplementary Figure 3). The energy relaxation rate $W_{ph}=2\times10^{14}$ s$^{-1}$ which is a fitting parameter. The scattering rate towards the LPB $W_{L0}$, a second fitting parameter, was taken $1\times 10^{14}$ s$^{-1}$ . $W_{U0}$ was taken equal to $W_{ph}$ (the injected particles have to relax down to LPB and decay from the cavity in less than 1 ps, because the Zeeman splitting corresponds to a full pseudospin rotation time of about 2.6 ps, and the observed angles are only a fraction of $2\pi$). The broadening for the LPB coupling was taken 12 meV by fitting the dependence on the detuning. The photonic SOC for the LPB and UPB were taken from the previous measurements ($T=15$ ps) \cite{Dufferwiel2017}. Exciton and photon lifetimes were taken equal to 5.3 and 15 ps \cite{Dufferwiel2017}.

The results of the simulations are shown in Fig. 5 of the main text.

\clearpage

\section{Supplementary Figure 1: Sample Image}

\begin{figure}[h]
	\includegraphics[scale=1]{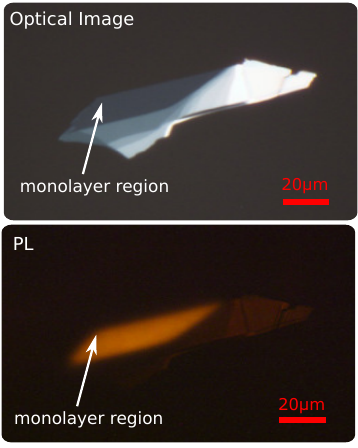}
	\caption{\label{fig:sample_image} \textbf{Top panel.} Microscope image of the monolayer region used in the measurements. \textbf{Bottom panel.} Photoluminescence map showing the optically active monolayer region of the flake which is around $20$x$60$ \micro m$^2$.}
\end{figure}

\clearpage

\section{Supplementary Figure 2: Exciton Valley Polarization}

\begin{figure}[h]
	\includegraphics[scale=1]{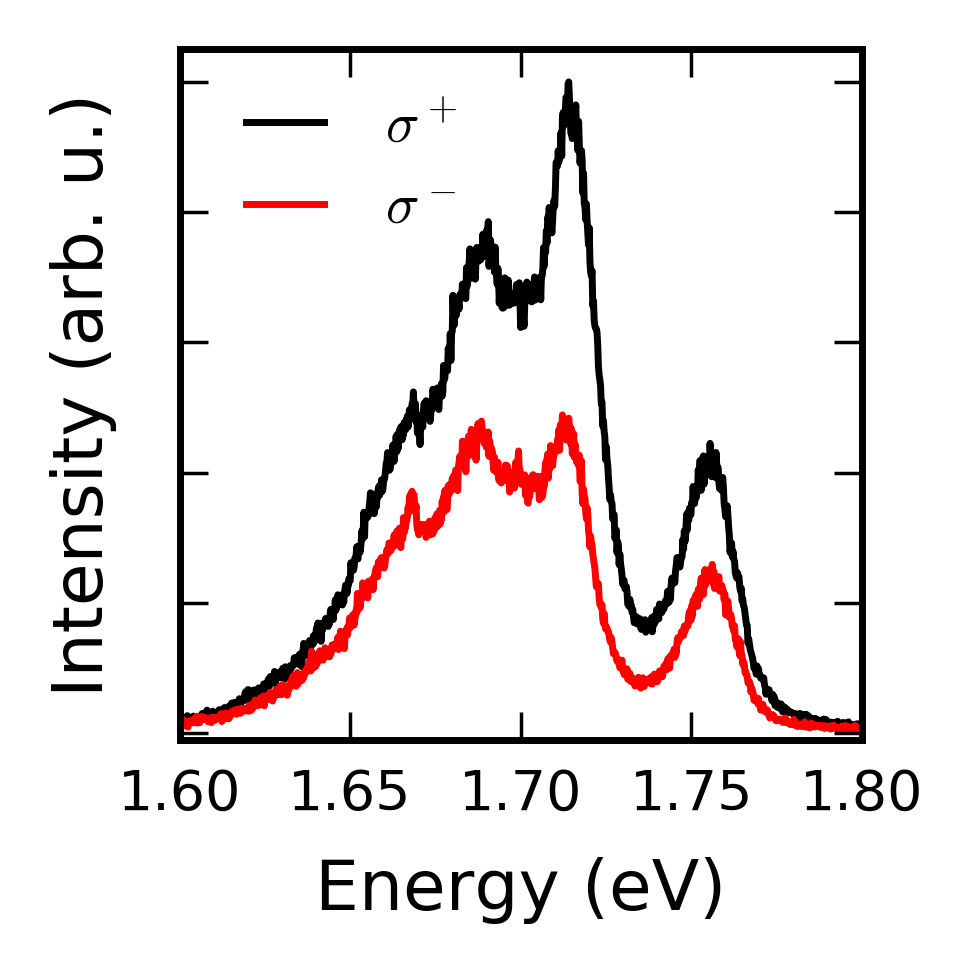}
	\caption{\label{fig:circ_pol} Exciton polarization resolved PL under $\sigma^+$ excitation.}
\end{figure}

Under circularly polarized excitation at $1.946$ eV clear retention of valley polarization is present for the excitonic species with circular polarization degrees of 30\% and 50\% for the exciton and trion respectively.

\clearpage

\section{Supplementary Figure 3:  Exciton valley Zeeman g-factor}

\begin{figure}[h]
	\includegraphics[scale=1]{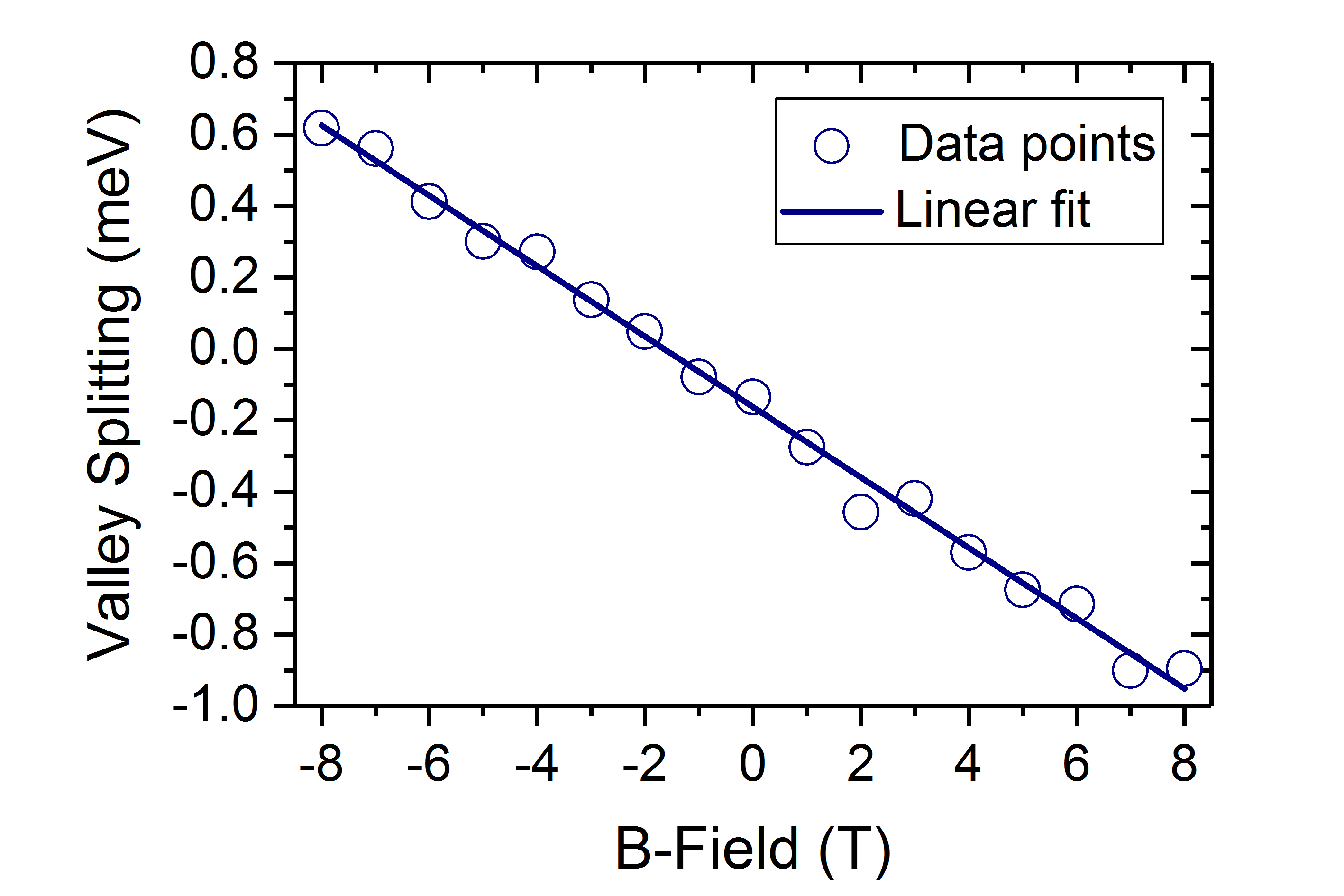}
	\caption{\label{fig:valley_zeemanl}Valley Zeeman splitting of the neutral exciton as a function of magnetic field. The fit corresponds to a g-factor of 1.7.}
\end{figure}

\clearpage

\section{Supplementary Figure 4:  Polariton Valley Polarization}

\begin{figure}[h]
	\includegraphics[scale=1]{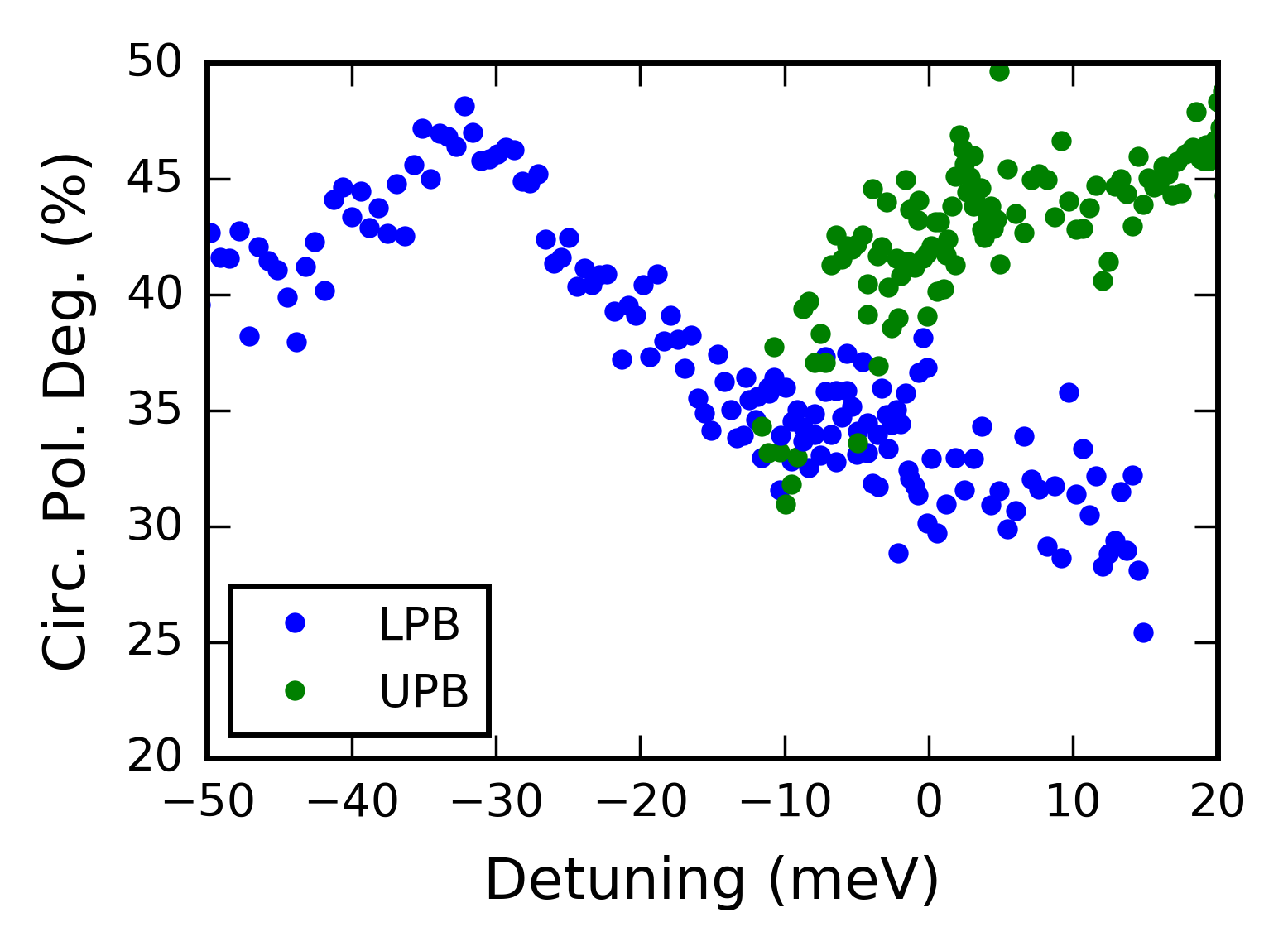}
	\caption{\label{fig:circ_pol} Circular polarization degree of the polariton branches under $\sigma^+$ excitation as a function of exciton-photon detuning.}
\end{figure}

The circular polarization degree of the polariton branches is plotted as a function of exciton photon detuning, $\Delta = E_c-E_{X^0}$, where $E_c$ and $E_{X^0}$ are the cavity  and exciton energies respectively, in Supplementary Fig.S4. We attribute the high polarization degree of the UPB to direct scattering of highly polarized high k-vector excitons, which have not significantly depolarized, and are degenerate with the UPB energy due to the parabolic exciton dispersion as discussed in detail in \cite{Dufferwiel2017} for MoSe$_2$ polaritons. The LPB has a polarization degree maximum when the weakly coupled mode is in resonance with the trion at around $\Delta = -30$ meV where the LPB polarization degree is comparable to that of the bare trion state. At positive detuning the LPB approaches the bare exciton polarization degree of around $30$\%. As discussed in \cite{Dufferwiel2017} the enhancement of the LPB polarization degree when slightly negatively detuned from the exciton resonance can be explained by taking into account cavity-modified relaxation of excitons, which relax quickly to the polariton branch before significant depolarization in the reservoir due to the exciton LT-splitting.

\clearpage

\section{Supplementary Figure 5:  LPB emission angle}

\begin{figure}[h]
	\includegraphics[scale=1]{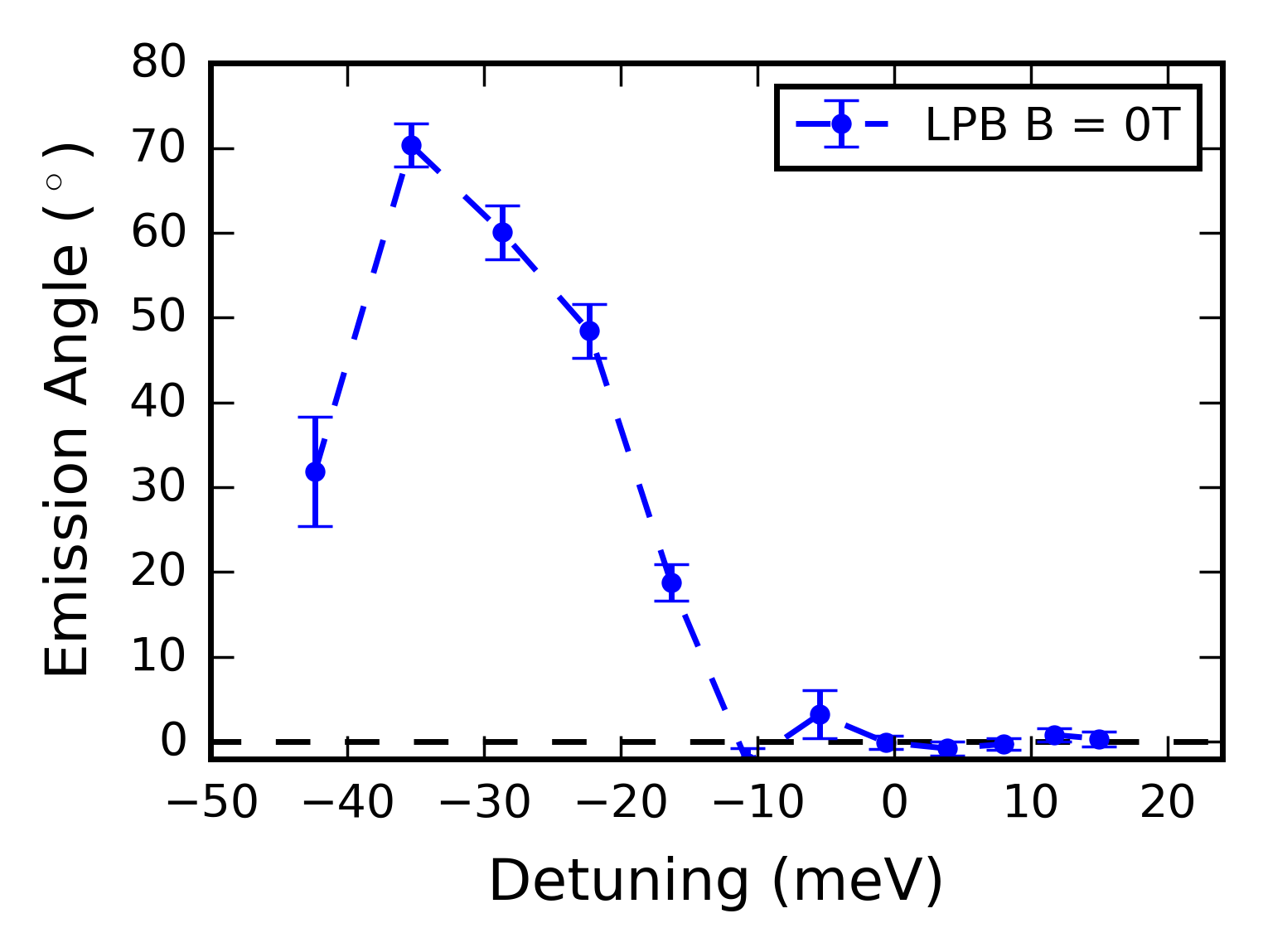}
	\caption{\label{fig:LPB_detuning} Emission angle of the LPB under vertical excitation as a function of detuning, at $B = 0T$.}
\end{figure}

Fig.~\ref{fig:LPB_detuning} shows the emission angle of the LPB under vertical excitation. When between detunings of $-10$ meV to $+20$ meV the linear emission angle of the LPB is $0^\circ$ which is defined by the vertical pump and indicates retention of injected coherence as discussed in the main text. Below $-10$ meV the LPB becomes quasi-resonant with the trion peak which is weakly coupled to the lower exciton-polariton branch. Here the LPB shows some slight rotation in the emission angle due to the polarization of the bare trion, which has an orthogonal polarization degree of $-3$\%, as shown in Fig.1\textbf{b} of the main text. This emission angle reaches a maximum when in resonance with the trion at a detuning of around $-30$ meV. As such, when measuring the coherent rotation due to the application of a magnetic field in Fig.4 of the main text, any rotation due to the applied field at detunings more negative than $-10$ meV are masked by the coupling of the trion and are not discussed.

\clearpage

%\bibliography{valley_coherence_wse2}

\end{document}